# Bending, Nanoindentation and Plasticity Noise in FCC single and poly crystals.

R. Bolin[b] , H. Yavas[a,d] , H. Song[a,b] , K. J. Hemker[a], S. Papanikolaou[a,b,c*]

[a] *Department of Mechanical Engineering, Johns Hopkins University, Baltimore, MD 21216*

[b] *Department of Mechanical and Aerospace Engineering, West Virginia University, Morgantown, WV 26506*

[c] *Department of Physics, West Virginia University, Morgantown, WV 26506*

[d] *Department of Control Engineering, Faculty of Electrical Engineering, Czech Technical University in Prague, Technika 2, Prague 6, Czech Republic*

*Corresponding Author: stefanos.papanikolaou@mail.wvu.edu

## 1. Abstract

We present a high-throughput nanoindentation study of *in-situ* bending effects on incipient plastic deformation behavior of polycrystalline and single-crystalline pure aluminum and pure copper at ultra-nano depths (<200nm). We find that hardness displays a statistically inverse dependence on in-plane stress for indentation depths smaller than *10nm*, and the dependence disappears for larger indentation depths. In addition, plastic noise in the nanoindentation force and displacement displays statistically robust noise features, independently of applied stresses. Our experimental results suggest the existence of a regime in FCC crystals where ultra-nano hardness is sensitive to residual applied stresses, but plasticity pop-in noise is insensitive to it.

## 2. Introduction

Nano-indentation provides a unique opportunity to probe mechanical deformation at the nanoscale of any solid surface. While numerous experimental nanoindentation studies have been conducted to understand nano and micro scale plasticity [1–18], it has been elusive to use surface nanoindentation to distinguish *surface* from *bulk* crystal plasticity features. In a dislocation-free region, nanoindentation turns from elastic to plastic through a sudden burst, labeled as primary "pop-ins" [19–28]. However, in a dislocation-rich region, nanoindentation is characterized by a noisy response, with multiple secondary pop-in bursts at multiple depths [29]. Nanoindentation primary pop-in bursts initiate crystal plasticity and are known to be driven by surface dislocation nucleation [30] due to large stress concentrations at the indentation tip. Surface-induced primary pop-in events have been very useful in revealing fundamental mechanisms of surface-induced crystal plasticity in annealed crystals, and they appear to provide an onset signature of local plasticity in dislocation-starved surface locations. In this work, we investigate the role and character of the secondary pop-ins, namely the collective noise that emerges during nanoindentation at low depths (<200nm) in determining mechanical properties of FCC crystals. We find that this secondary noise is robust across FCC single and poly- crystals, as well as across applied in-situ tension, even though hardness displays strong sensitivity.

Nanoindentation has been used in investigations of *in-situ* mechanically loaded metallic samples using a variety of indentation protocols, geometries and depths [31-37]. The main outcome of these works is a strong dependence of the apparent hardness on tensile stress, which nevertheless disappears if the nanoindentation surface contact area is scaled appropriately [38, 39]. However, nanoscale effects of pop-in events, either primary or secondary ones, and especially in *in-situ* loaded samples with large pre-existing dislocation density [30,31] (namely, pre-strained and *in-situ* loaded) have been ambiguous [35]. In particular, the question that rises is the ability to distinguish bulk dislocation density populations through the thorough investigation of noise features in force and hardness measurements at the ultra-nanoscale regime (<200nm).

In this paper, we focus on the statistical features of the noise in the nanoindentation force-depth curves at very shallow depths (<200nm). We concentrate our efforts on common FCC metals, in particular single and poly crystalline pure aluminum, and single crystal pure copper. In addition, we explore the effect of in-situ bending stress on nanoindentation at very shallow depths. At shallow depths, plasticity is not primarily controlled by the shape of the indenter tip and the most pronounced evidence of this fact is the well known observation that the post-indentation surface profile is stochastic at these depths and does not *exactly* follow the indenter tip's shape [30,31]. We study two tips, Berkovich and spherical with radius 5μm, and we find qualitative agreement. In this work, we report on high-throughput indentation measurements on *pre-stressed* FCC samples and statistically analyze large datasets of load-displacement curves, focusing on the behavior of hardness and pop-in noise. High-throughput indentations may be averaged to statistically nullify the effects of *uncorrelated* surface roughness and/or grain orientation, thus providing us the opportunity of only focusing on intrinsic microstructural effects. Single crystalline samples were oriented at {100} orientation, since uniaxial tension along this orientation leads to multi-slip dislocation plasticity [40]. The material selection of Cu and Al provided also some testing on the variability of stacking fault energies and propensity for cross-slip in two distinct microstructures [41,42].

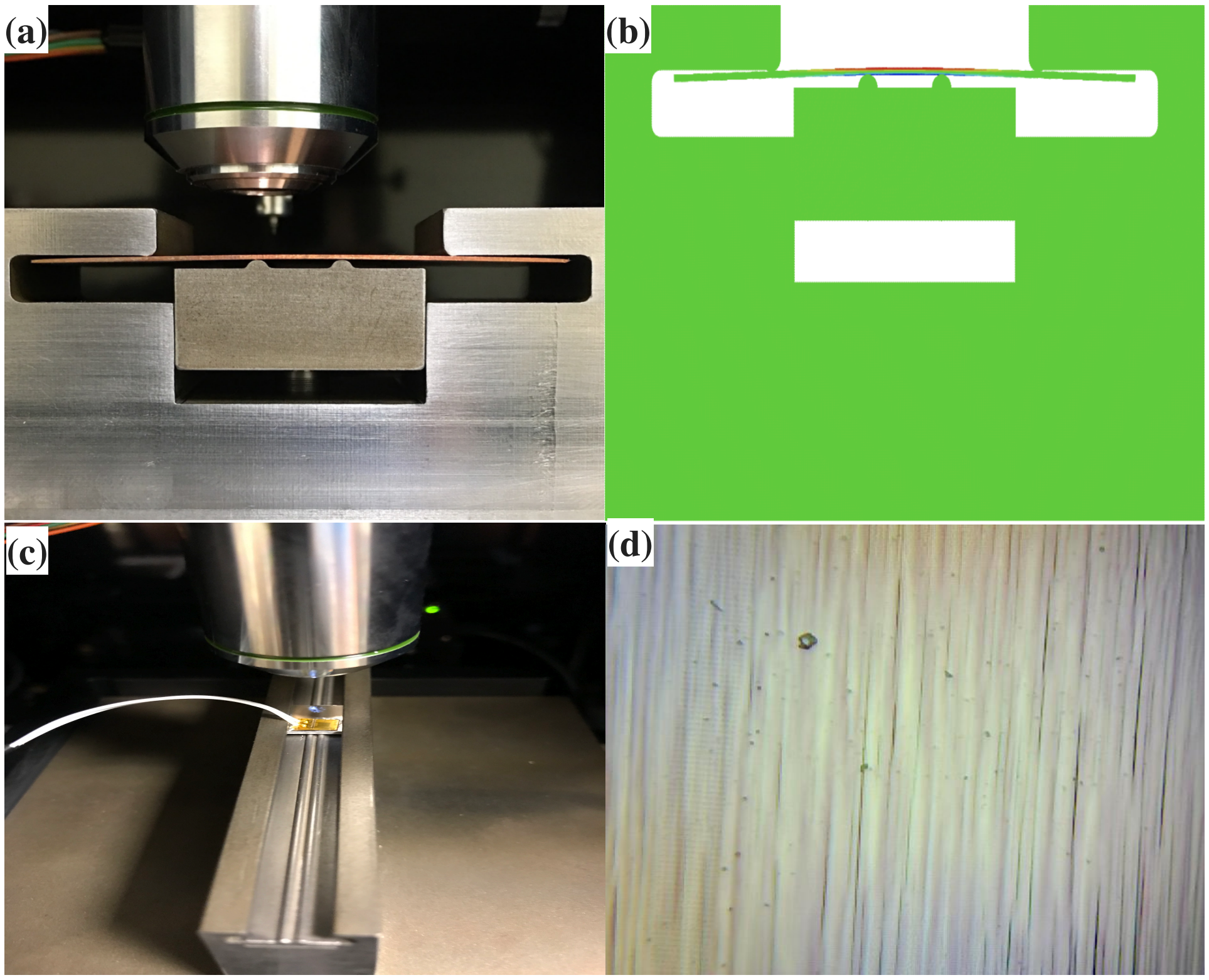

**Figure 1: Representative images of experimental device** (a) custom 4pt-bending fixture, side view of the stage for poly-crystalline sample, (b) finite element modeling, using ABAQUS, assisted design, (c), top view of 4pt-bending fixture for single-crystalline sample, strain gauge is glued on the top surface of the sample. (d) optical image of surface steps when strain is larger than 0.1% in single crystalline copper sample.

## 3. Experimental Details

Electropolished commercial aluminum polycrystals (99.99% purity, Plasma Materials Inc., US) and commercial aluminum and copper single crystals at orientation (100) (99.99% purity, MTI Corporation., US), of dimensions 2mm x 5mm x 10mm, were used in this study. Custom-made 4-point loading fixtures (see Figure 1(a),(c)) were designed to apply in-plane tension at the sample top central region during nanoindentation, controlling the local strain at the top central surface area through a screw element at the bottom of the fixture. The strain was measured *in situ* by using commercial strain gauges (see Fig.1c). High-throughput nanoindentations were performed in the center $(1mm)^2$ surface areas of the samples, with typical distances between nanoindentation sites being 10μm in each direction. Given that nanoindentation depths did not exceed 250nm, the distance between nanoindentation sites may be regarded independent [35]. For the estimation of the applied tension at the nanoindentation sites, independently measured elastic moduli and yield stresses were exported into finite element simulations, performed using the ABAQUS software (see an example in Figure (1)b and also in the Supplementary Material (SM)). Through systematic calculations and testing, tables for strain/stress/plastic-strain mapping were developed for each material (see Tables 1,2 for a particular example of polycrystalline, as well as single-crystalline aluminum). The applied strain on the samples extended, in small steps, up to 0.5%, well in the crystal plasticity regime. Clear surface steps (primarily due to dislocation plasticity) formed after 0.2% strain in all the materials tested (example seen in Fig.1d for a single crystal copper sample), naturally influencing the indentation results at small depths. Our main results are focused on small loads/strains which should not be influenced by such steps, but more details on these issues are discussed in the SM.

| Deflection, μm | Total strain, % | Plastic strain, % | Tensile stress, MPa |
|---|---|---|---|
| 0 | 0 | 0 | 0 |
| 97 | 0.031 | 0 | 19.1 |
| 226 | 0.065 | 0 | 40.13 |
| 316 | 0.11 | 0.04 | 41.86 |
| 403 | 0.15 | 0.08 | 43.21 |
| 493 | 0.2 | 0.127 | 44.6 |
| 583 | 0.24 | 0.165 | 45.9 |
| 717 | 0.31 | 0.233 | 47.12 |
| 823 | 0.36 | 0.283 | 47.35 |

**Table 1: Example of Bending deflection-Strain-Stress Correspondence:** Polycrystal *Al*: Measured total strain and calculated stress and plastic strain.

| Deflection, μm | Total strain, % | Plastic strain, % | Tensile stress,MPa |
|---|---|---|---|
| 2.3 | 0.005 | 0 | 3.5 |
| 9.3 | 0.05 | 0.009 | 20 |
| 11.2 | 0.1 | 0.04 | 20.11 |

**Table 2: Example of Bending deflection-Strain-Stress Correspondence:** Single Crystal *Al*: Measured total strain and calculated stress and plastic strain.

Nanoindentation experiments were performed with an iNano (Nanomechanics Inc., TN) nanoindenter with Berkovich (apex roundness of 20nm) and spherical (5μm) tips, acquired by Microstar Inc.. The details of the materials preparation, bending fixture and nanoindentation protocols are discussed in the SM.

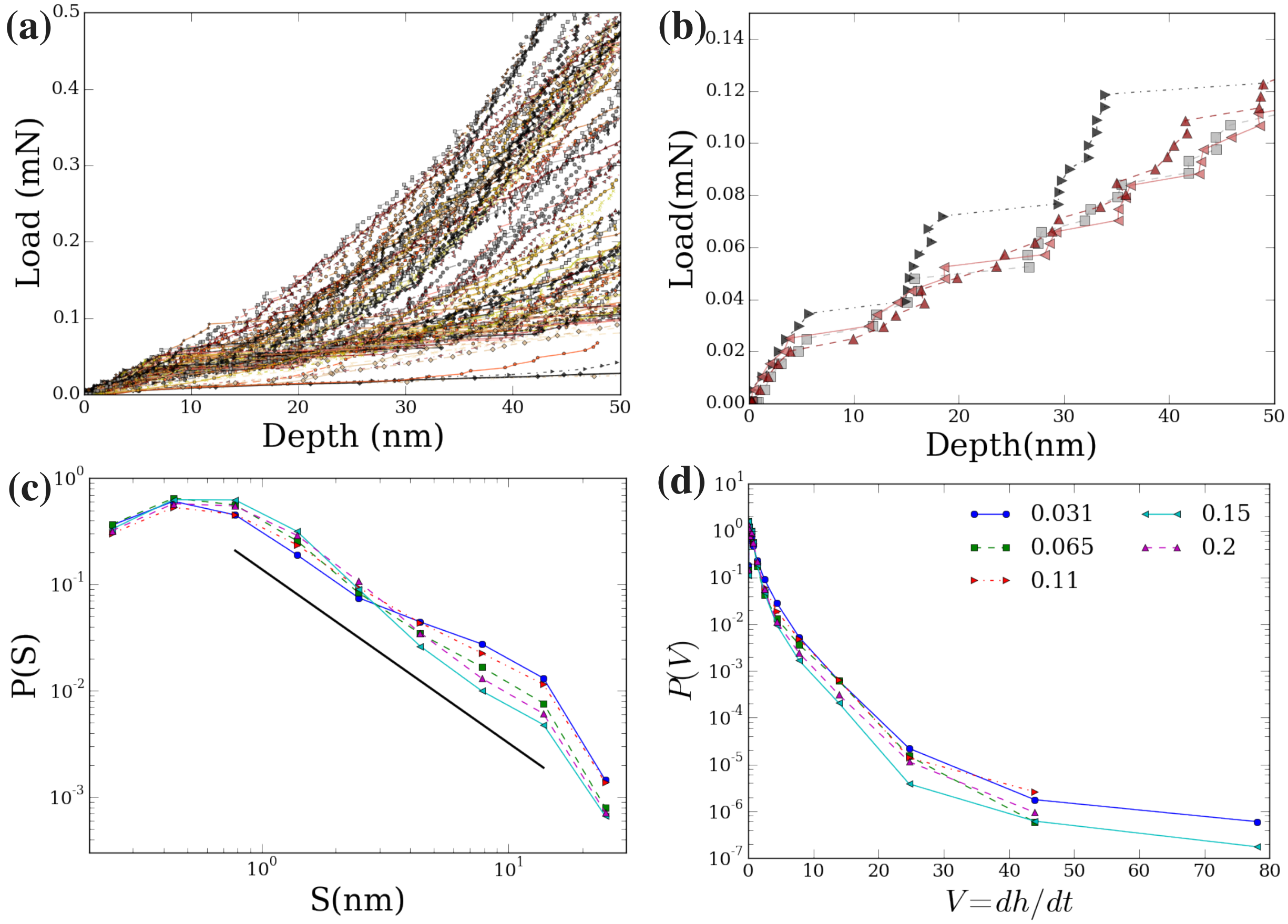

**Figure 2: Pop-in Events in polycrystalline aluminum** (a) A representative sample of nanoindentation load-depth curves in Al non-stressed samples over a region of $1mm^2$ on the top surface, (b) Detailed load-depth behavior at four randomly selected locations on non-stressed sample top surface, (c) Probability event distribution P(S) as function of event size S (described in text) for applied total in-plane strain of 0.031 (●) (blue), 0.065 ▌ (green), 0.11 (▶) (red), 0.15 (◀)(cyan), 0.2 (▲)% (purple). The solid line is a guide to the eye and represents y ~ $x^{-1.6}$ (d) Probability distribution P (V ) of the "pop-in noise" intensity V = dh/dt (described in text) (total strain shown in legend).)

## 4. Results

In the following, we present our main results on the correlation between hardness and secondary pop-in bursts in FCC polycrystalline pure aluminum, single-crystalline pure aluminum and single crystalline pure copper. Our data is also supported by the SM which provides additional details. In each of the cases, we focused on two main observables: i) the raw force-depth curves and ii) the Continuous Stiffness Method (CSM)[35]-estimated hardness. Both observables were analyzed in statistical terms, by defining histograms of hardness or large fluctuations in depth, as the applied tension is modified. For the pop-in event probability distributions, the pop-in noise event size is defined so that $S = \sum_{i for \delta h > h_{thr}} \delta h_i$ , to analyze the displacement bursts in a quantitative manner, using the threshold $h_{\text{thr}}$ being equal to the machine noise threshold $h_{\text{thr}}$=0.2$nm$, which was independently tested, given the laboratory's environmental conditions.

<u>Polycrystalline Al:</u> In order to investigate the influence of the applied in-plane stresses on the load-displacement curves, we carried out indentation tests across a large surface area of 1x1 mm$^2$ in the centerline of three polycrystalline specimens (for each sample ~5,000 indentations at each stress level given in Table 1). Samples were electropolished before being loaded and indented.

In all cases, load-displacement curves show a continuous elastic response followed by multiple measurable displacement bursts (see Figure 2(b)). These bursts are at depths large compared to the expected oxide film in aluminum, and therefore may statistically attributed to the mechanical response of polycrystalline aluminum. This intrinsic material noise is attributed to both surface roughness and stochasticity of crystalline plastic deformation. In Figure 2(c), $S$ is the magnitude of a single displacement burst, while $P(S)$ is the probability density. The event distributions display a remarkable stability into a form that resembles P(S)=A * $S^{-1.6}$ exp(-S/$S_0$), where A is a normalization constant and $S_0$ defines the cutoff of the distribution (here, ~12nm in all applied tensions). Interestingly, the applied tension appears to consistently suppress large events, however the robustness of the power-law response ($P(S) \sim S^{-1.6}$) seems clear. It is worth noting that this power-law response is observed in other nanomechanical studies [7,43] , and it is suspected to hold across nanomechanical responses in a material-independent and loading-independent, thus universal, manner [41,44,45,46]. The current study represents the first evidence of this kind in nanoindentation of FCC crystals. Following Ref.[45] and in the effort to corroborate the evidence of Figure 2(c), we also investigated the behavior of the local event "intensity", defined as the recorded rate of depth changes dh/dt. In Figure 2(d), the probability distribution of the local event intensity $P(V)$ is presented as obtained data from depth vs. time ($h-t$) curves. The observed behavior is reminiscent of typical avalanche dynamics in various mean-field models [45] and appears independent of the applied tension.

A distinct dimension in the investigation of nanoindentation is provided by the behavior of hardness. While not an exact measurement, the CSM method [35] provides a concrete continuous measure of hardness that is expected to be consistent in relative terms, especially for the same material class and sample. Figures 3(a) to 3(d) show the variation of hardness as a function of indentation depth at different in-plane stresses, for a large multitude of indented locations and five different samples. Hardness displays a strong size effect dependence on depth for all in-situ tension levels. These size effects are well known from various prior studies of Berkovich indentation on FCC metals [27,30,35 and references therein]. Hardness values first increase with increasing indentation depth until it reaches a peak (approximately 2 GPa for zero applied in-plane stress and zero strain, see Fig. 2c and 2d) and then decreases towards a plateau at approximately 1 GPa consistent with prior studies [22,24,26,36,37]. However, we notice an additional, overall, unseen before, effect of *in-situ* tension on hardness, which becomes clear when Figure 3(a) (low tension) is contrasted to Figure 3(b) (high tension). Despite natural data variability due to extrinsic (roughness) and intrinsic (dislocation bursts) reasons, there is a strikingly strong correlation between the depth-dependent hardness and the applied in-plane stress. At up to 43.21 MPa in-plane stresses, the hardness shifts towards lower values. As it can be seen in Figures 3(b), 3(c), 3(d), the hardness below depths of 10 *nm* shows a clear dependence on the applied in-plane stress (strain), while for larger depths such dependence disappears.

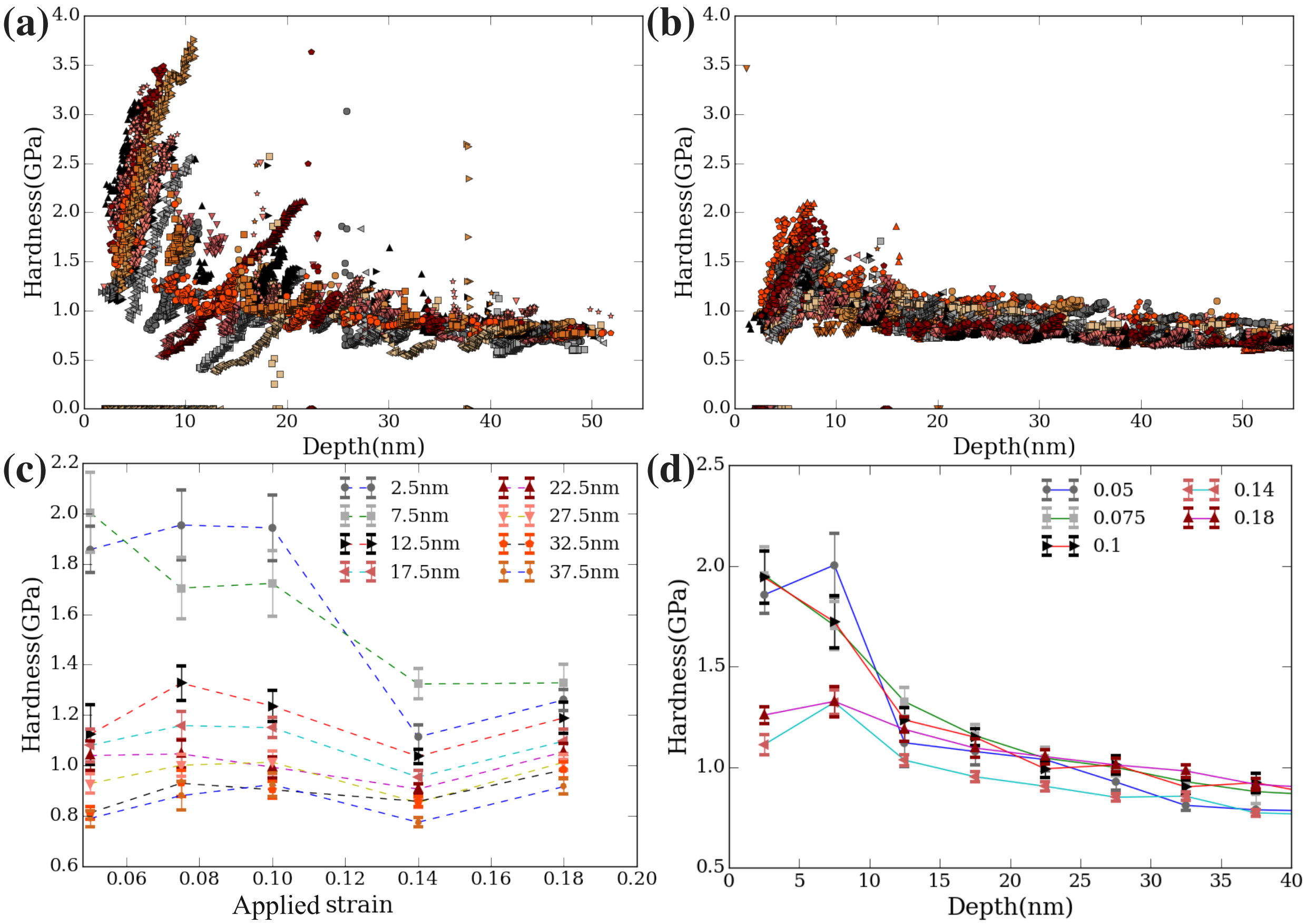

**Figure 3: Hardness in poly-crystalline Aluminum:** Variation of hardness as a function of indentation depth for multiple samples at in-plane stress (a) zero and (b) 43.21 MPa (with 0.08% plastic strain (see Table 1)). In (c) we show the change of average and binned hardness with applied in-plane strain at various indentation depths (depths shown in legend), while in (d) we show the change of average and binned hardness with depth at various total applied strains (corresponding strains shown in legend).

*Single-crystalline Al:* Samples at orientation (100), of ultra-high purity, were mechanically polished to nanometer scale as purchased from manufacturer. Analogously to the case of polycrystalline *Al*, Figures 3 (a) and (b), (c) show force-displacement curves and the statistics of pop-in events, for different applied tension levels. Figure 3(a) displays the actual response of four individual indentation sites, demonstrating the existence of non-trivial displacement jumps that extend to 10nm, even at depths of 100nm. If these curves are averaged across indentation sites, then the average dependence on the applied tension can be distinguished at very low depths (<20nm) as well as large depths (>125nm). Nevertheless, the histograms of displacement bursts through all indentation sites at a given stress level display a wide distribution that has a marked independence on the applied tension. While the power-law behavior appears similar to the one observed in polycrystalline *Al*, the behavior resembles $P(S)= A^* S^{-1.3} \exp(-(S/S_0)^{1.5})$. These differences may signify distinct micromechanical mechanisms, as expected [45,46].

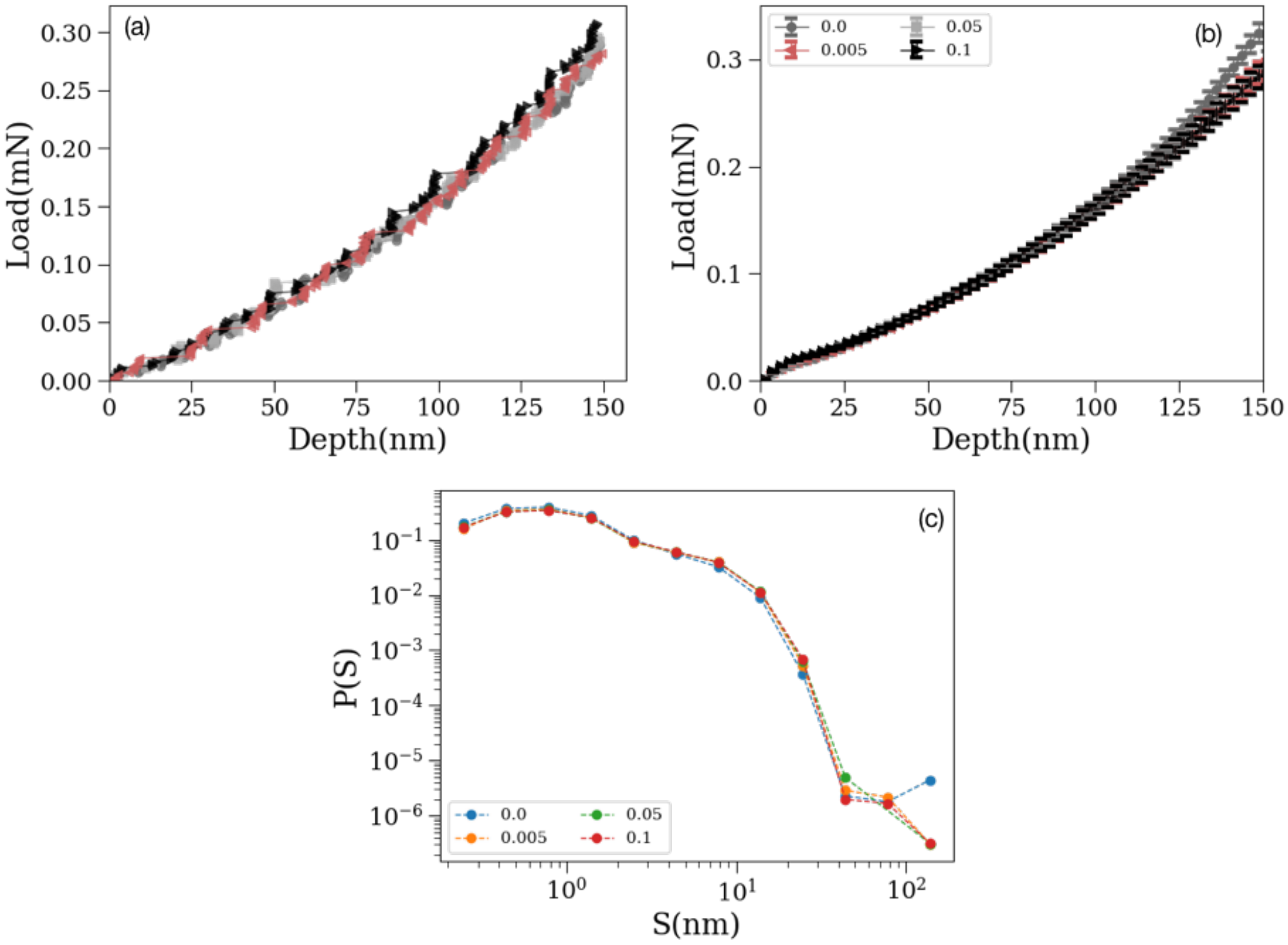

**Figure 4: Pop-in Events in single crystal aluminum** (a) Representative samples of nanoindentation load-depth curves in Al non-stressed samples, (b) Sample-averaged load-depth behavior as function of applied total tensile strain levels (0%,0.005%,0.05%,0.1%). There are about 1000 tests averaged per strain. (c) Probability event distribution P(S) as function of event size as function of the applied in-plane tensile strain.

The behavior of the CSM hardness for single crystalline *Al* is quite consistent with the polycrystalline case. As seen in Figure 5(a), the sample-averaged hardness shows that the applied tension drastically decreases the small-depth hardness by a factor of 30. As shown in Figure 5(b), the decrease of the hardness at depths less than 15nm is proportional to the applied strain at a high negative power n<-3. Figure 3(d) shows the sample-average hardness at various depths as function of applied tensile strain, showing both the saturation at larger strains and the drastic decrease as soon as tension is applied. AFM analysis and indentation arrays of the single crystalline films signify the indentation site independence (see SM for details).

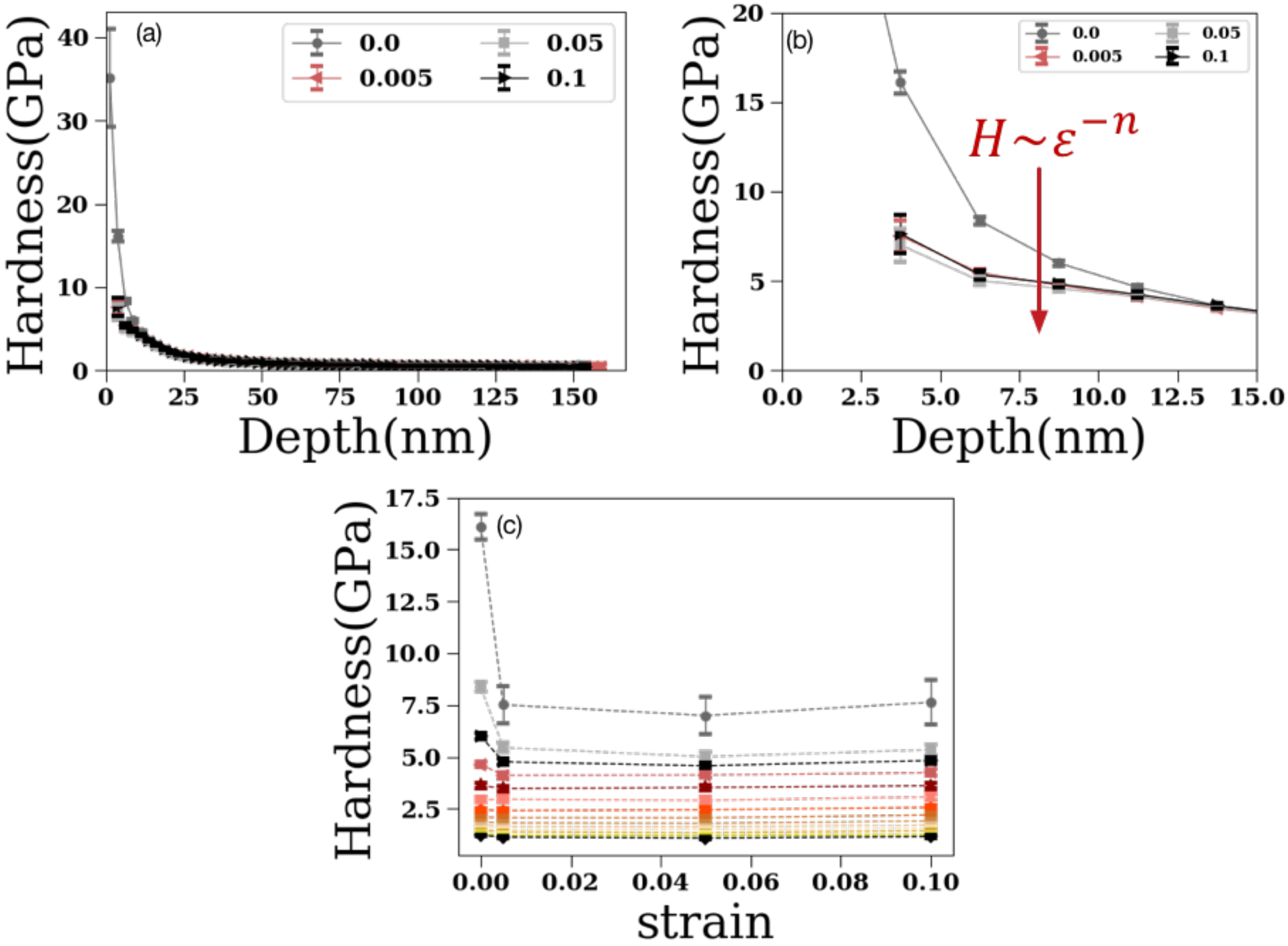

**Figure 5: Hardness & Statistics in Single-Crystalline Aluminum:** (a) Sample-averaged hardness-depth measurements carried on single crystals (100) at multiple tensile strain levels (0%, 0.005%, 0.05%, 0.1%). (b) Average hardness for 4 strains (zoomed in) showing that the behavior is consistent with a strong dependence on applied tensile strain, analogously to the behavior in polycrystalline *Al*, showing that hardness is proportional to a power law of the applied tensile strain. Different symbols/colors indicate different applied tensile strains. Each figure has almost 1000 indentations averaged in each strain. (c) shows the full hardness averages with respect to the applied tensile strains, for a given depth bin.

*Single-crystalline Cu:* Samples at orientation (100), of ultra-high purity, were mechanically polished to nanometer scale as purchased from manufacturer. Analogously to the other materials tested, results display a universal set of displacement bursts at small depths in conjunction to a strong sensitivity of the CSM hardness on *in-situ* applied tension. Figure 6(a) shows sample responses at individual indentation sites. Displacement bursts are clear but evidently smaller than in Aluminum. The sample-averaged force-displacement curves in Figure 6(b) show a clear dependence on the applied tension at all tested depths. In Figure 6(c), the histogram of displacement bursts show the existence of a clear power-law behavior that resembles the one found for polycrystalline aluminum following the form $P(S)=A * S^{-1.6} \exp(-S/S_0)$.

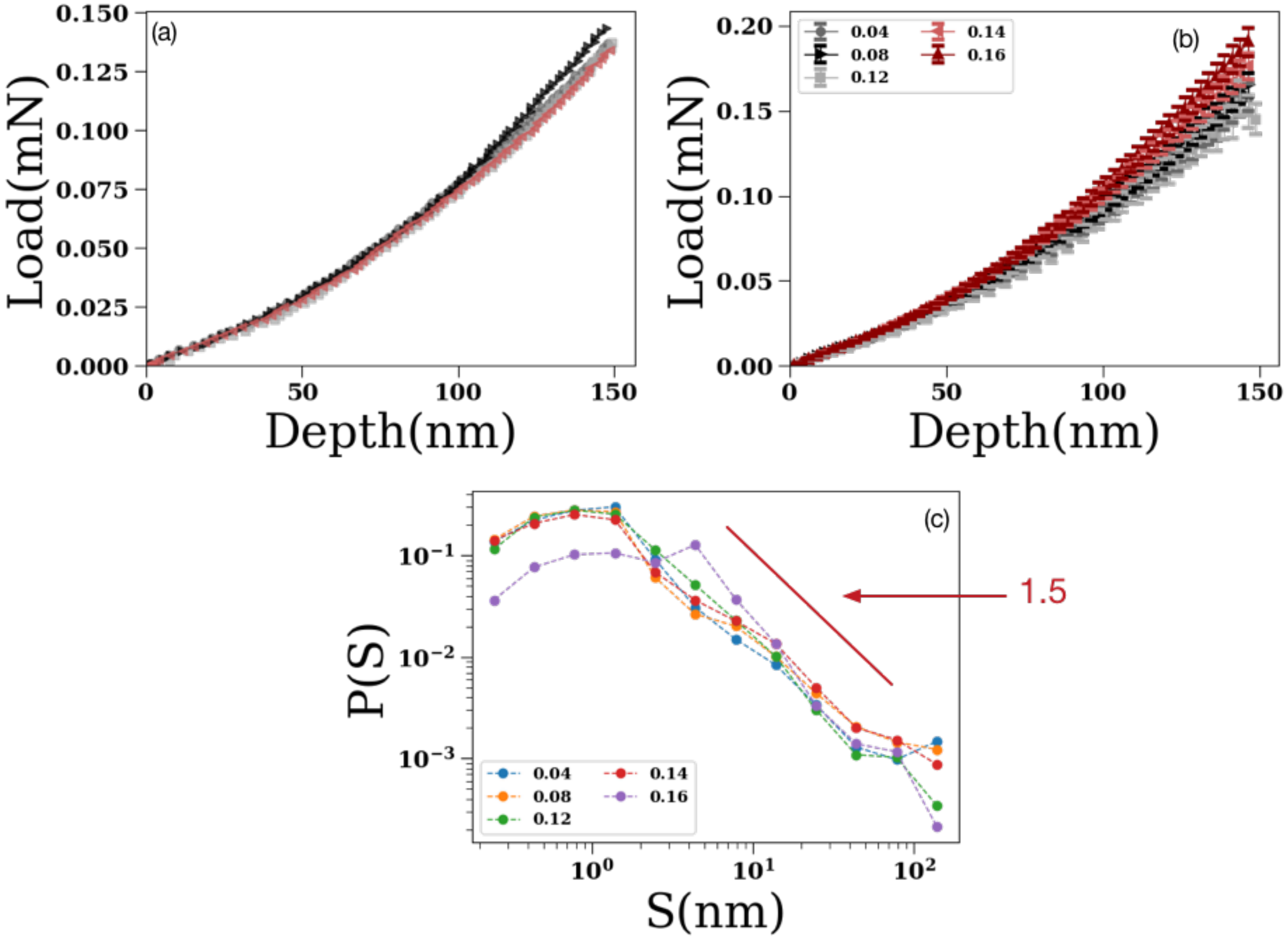

**Figure 6: Force-Displacement and Pop-in Events in single crystal copper** (a) Representative samples of nanoindentation load-depth curves in *Cu* non-stressed samples, (b) Sample-averaged load-depth behavior as function of applied total tensile strain levels (0.04%,0.08%,0.12%,0.14%,0.16%). There are about 1000 tests averaged per strain. (c) Probability event distribution P(S) as function of event size as function of the applied in-plane tensile strain. The statistical behavior is similar to the one observed for poly crystal copper (not shown [47]), and consistently smilar to the behaviors observed for single crystal and poly crystal *Al*.

The behavior of the CSM hardness for single crystalline *Cu* is quite consistent with the polycrystalline case. As seen in Figure 5(a), the sample-averaged hardness shows that the applied tension drastically decreases the small-depth hardness by a factor of 30. As shown in Figure 5(b), the decrease of the hardness at depths less than 15nm is proportional to the applied strain at a high negative power n<-2. Figure 3(d) shows the sample-average hardness at various depths as function of applied tensile strain, showing again, both the saturation at larger strains and the drastic decrease as soon as tension is applied.

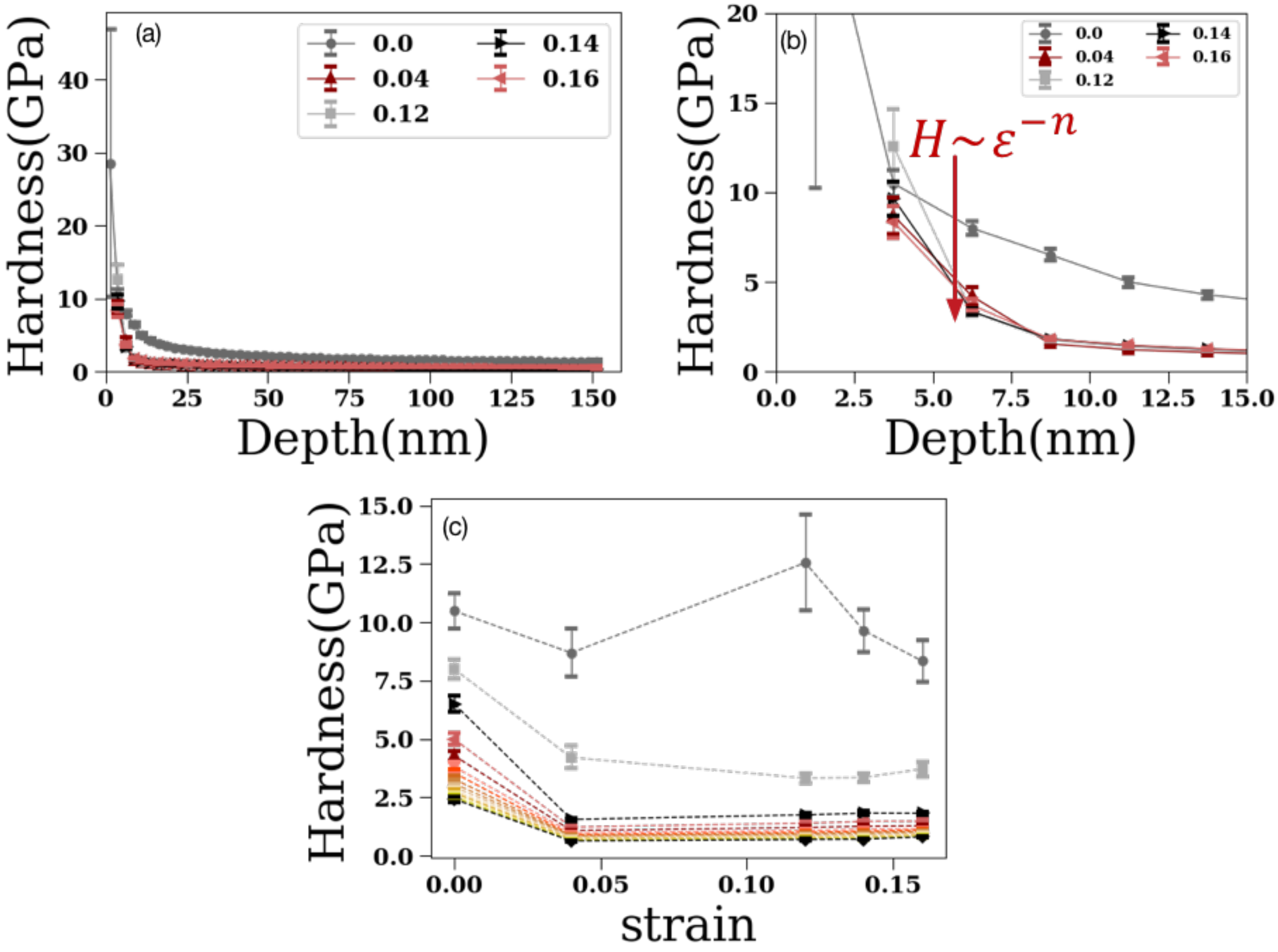

**Figure 7: Hardness & Statistics in single-crystalline copper:** (a) Sample-averaged hardness-depth measurements carried on single crystals (100) at multiple tensile strain levels (0%, 0.04%, 0.08%, 0.16%). (b) Average hardness for 4 strains (zoomed in) showing that the behavior is consistent with a strong dependence on applied tensile strain, analogously to the behavior in single and poly crystalline *Al*, showing that hardness is proportional to a power law of the applied tensile strain. Different symbols/colors indicate different applied tensile strains. Each figure has almost 1000 indentations averaged in each strain. (c) shows the full hardness averages with respect to the applied tensile strains, for a given depth bin. (a-c) shows that the hardness at shallow depths transitions almost immediately, but the effect is still present till 15nm.

## 5. Discussion and Conclusions

The variation of depth-dependent hardness at the ultra-nano regime has remained a challenging concept: Early studies [22,24,36,48] used the strain gradient plasticity approaches using geometrically necessary dislocations concepts. However, further experimental investigations were not convincing enough to support the overall conclusions, particularly at nanometer-scale depths [37]. Specifically, the Nix-Gao model overestimates the hardness values at very small depths [48]. Poole et al. suggested that the overestimation is due to the strong repulsion of the geometrically necessary dislocations (GND) at shallow depths, causing an expansion of the effective volume of geometrically necessary dislocations [49,50]. Similarly, Nix and Feng reported similar observations in later studies [51]. According to their study, a significant number of dislocations that spreads out of the plastic zone causes relaxation. In brief, the nanoscale plausible plasticity is governed by two main mechanisms, either dislocation nucleation at the surface or interactions, movement and nucleation of the pre-existing bulk (but close to the surface) dislocations. While the former case has been well known, the precise role of the latter effect is the focus of this work. In order to understand the effect of bulk dislocation interactions and nucleation, we decided to perform simulations in the absence of surface dislocation nucleation – in this way, the overall qualitative effect of bulk dislocation driving is explored and understood.

To explore possible physical phenomena that could lead to the observed hardness deviation, we have used two-dimensional discrete dislocation dynamics (2D DDD) simulations [51] (see also SM) to account for the effect of in-plane stress and indentation depth. We assumed that bulk dislocation sources are prevalent and also, indentation depths are small so that indentation is primarily dominated by close-to-surface but bulk dislocation nucleation and movement. In this way, we did not explicitly model surface dislocation nucleation. Details of the 2D DDD model can be found in the SM and in Ref.[51]. The density of pre-existing dislocations was set at $3\times10^{12} m^2$. We noticed that for depths smaller than $10nm$, and holding the depth fixed, the increase of in-plane stress results in a sharp decrease of the indentation force. This result is consistent with the experimental findings reported in this work.

Furthermore, one may apply small-scale plasticity considerations using a local yield stress picture framework. Gerberich et al. [52] linked the indentation size effect in the nanometer scale to a ratio between the energy of newly created surface and plastic strain energy dissipation. The hardness in the nanometer scale then follows, $H \simeq \frac{\sigma_f}{\left(\frac{S}{V}\right)^{2/3}} \frac{1}{(2\delta R)^{1/3}}$ for spherical indentation (tip radius $R$) where $S/V$ is the plastic surface area over volume ratio, $\delta$ is indentation depth. Based on the data of Au in table 2 of [52], $\frac{S}{V}\sqrt{\delta}$ decreases at small indentation depth. We define $\sigma_f$ as the local material flow stress, which is a function of the local dislocation density. In nanoindentation, the local flow stress should determine the measured hardness. The local flow stress is expected to[53] have a complex non-monotonic dependence on the local dislocation density at the nanoscale, in analogy with early theoretical suggestions as well as studies of metallic nanopillars [54]. Following Ref. [54], we suggest that the local flow stress during indentation is a function of the local dislocation density: $\sigma_f \propto \frac{\beta}{R\sqrt{\varrho}} + \alpha b\sqrt{\varrho}$, where $\varrho$ is the local dislocation density, $R$ is the nanoindenter's radius, $b$ is the magnitude of the Burgers vector, and $\beta$, $\alpha$ are dimensionless fitting parameters which in our case take the values $1.76\times10^{-3}$ and 0.46, respectively. For the indentation depth of 5 *nm*, our estimate of hardness vs. in-plane stress is shown in Fig. 4b (dashed line), which qualitatively agrees with our simulation and experimental findings.

In the light of our nanoindentation experiments and their agreement with the simple DDD simulations [51], we propose an alternative explanation for hardness size effects at the ultra-nano regime [29]: for small indentation depths (below 10 *nm* for a Berkovich tip, but more generally below the overall depth at which pop-ins would be expected), the applied stress-induced dislocation motion *statistically* dominates the deformation, in the sense of ensemble-averaged behavior over many indentation locations. At that length-scale, initial dislocation density controls the deformation behavior of the sample. This behavior is analogous to the source-limited regime found in pillar compression [43,44,54]. For indentation depths above 10*nm*, the dislocation density saturates, and the system reaches the critical GND density threshold and is independent of the applied in-plane stress. As the indentation depth increases, the dislocation density is controlled by dislocation source nucleation, and the effect of dislocations generated by in-plane tension disappears at large indentation depth (>50 *nm*).

In summary, we employed large arrays of nanoindentation tests on polycrystalline and single crystalline pure FCC metals (Cu, Al) at different in-plane tensions to investigate the incipient plasticity transition and size effect dependences. The depth-dependent hardness measurements show a clear transition at ~10 *nm*, as the applied in-plane stress increased to ~50*MPa* and the estimated in-plain strain rose to 0.3%. That is indicative of the high stochastic behavior as small indentation depths disappeared at high in-plane stresses, while pop-in statistics indicate that displacement bursts are insensitive to in-plane tension. The experiments are comparable to 2D DDD simulations and a plausible constitutive dislocation density model.

## 6. Acknowledgments

We would like to thank Bryan Crawford at JHU for technical support throughout this work. This research was supported by the U.S. Department of Energy, Office of Sciences, Basic Energy Sciences, DE-SC0014109 and also, the National Science Foundation, Award #1709568.